\documentclass[fleqn,usenatbib]{mnras}

\usepackage{mathptmx}

\usepackage[T1]{fontenc}

\DeclareRobustCommand{\VAN}[3]{#2}
\let\VANthebibliography\thebibliography
\def\thebibliography{\DeclareRobustCommand{\VAN}[3]{##3}\VANthebibliography}

\usepackage{graphicx}	
\usepackage{amsmath}	
\usepackage{bm}

\usepackage{txfonts}
\usepackage{xcolor}
\usepackage{natbib}
\usepackage[utf8]{inputenc}

\title[Lorentz force mediation of turbulent dynamo transitions]{Lorentz force mediation of turbulent dynamo transitions}

\author[K. M. Soderlund et al.]{
Krista M. Soderlund,$^{1}$\thanks{E-mail: krista@ig.utexas.edu}
Paula Wulff$^{2}$, Petri J. K{\"a}pyl{\"a},$^{3}$, 
and Jonathan M. Aurnou$^{2}$
\\
$^{1}$Institute for Geophysics, Jackson School of Geosciences, The University of Texas at Austin, USA\\
$^{2}$Department of Earth, Planetary, and Space Sciences, University of California, Los Angeles, USA\\
$^{3}$Institute for Solar Physics (KIS), Georges-K\"ohler-Allee 401a,
Freiburg, 79110, Germany\\
}

\date{Accepted XXX. Received YYY; in original form ZZZ}

\pubyear{2025}

\begin{document}
\label{firstpage}
\pagerange{\pageref{firstpage}--\pageref{lastpage}}
\maketitle

\begin{abstract}
We investigate how the strength of the Lorentz force alters stellar convection zone dynamics in a suite of buoyancy-dominated, three-dimensional, spherical shell convective dynamo models. This is done by varying only the magnetic Prandtl number, $Pm$, the non-dimensional form of the fluid's electrical conductivity $\sigma$. Because the strength of the dynamo magnetic field and the Lorentz force scale with $Pm$, it is found that the fluid motions, the pattern of convective heat transfer, and the mode of dynamo generation all differ across the $0.25 \leq Pm \leq 10$ range investigated here. For example, we show that strong magnetohydrodynamic effects cause a fundamental change in the surface zonal flows: differential rotation switches from solar-like (prograde equatorial zonal flow) for larger electrical conductivities to an anti-solar differential rotation (retrograde equatorial zonal flow) at lower electrical conductivities. This study shows that the value of the bulk electrical conductivity is important not only for sustaining dynamo action, but can also drive first-order changes in the characteristics of the magnetic, velocity, and temperature fields. It is also associated with the relative strength of the Lorentz force in the system as measured by the local magnetic Rossby number, $Ro_\ell^M$, which we show is crucial in setting the characteristics of the large-scale convection regime that generates those dynamo fields.
\end{abstract}




\section{Introduction}

The dynamics of stellar convection zones are dominated by nonlinear interactions between large-scale differential azimuthal flows and meridional overturning circulations, which together drive the generation of stellar-scale magnetic fields \cite[e.g.,][]{MieschToomre09,kapyla2023simulations}. The interplay of turbulent convection and large-scale circulations also acts to drive latitudinal gradients in surface luminosity, which can act back on the differential rotation 
\citep[e.g.][]{AurnouEA08, SoderlundEA13, KapylaEA2020}. Such complex systems can have multiple behavioral regimes \citep[e.g][]{VivianiEA19, HindmanEA20, Menu_20, Camisassa22, ZaireEA22}. Defining these regimes and elucidating their essential physics are necessary to understand the behavior of our sun and stars in general \citep[e.g.][]{Reinhold15, MetcalfeEA2016,LehtinenEA21}.

\begin{figure}
\includegraphics[width=\linewidth]{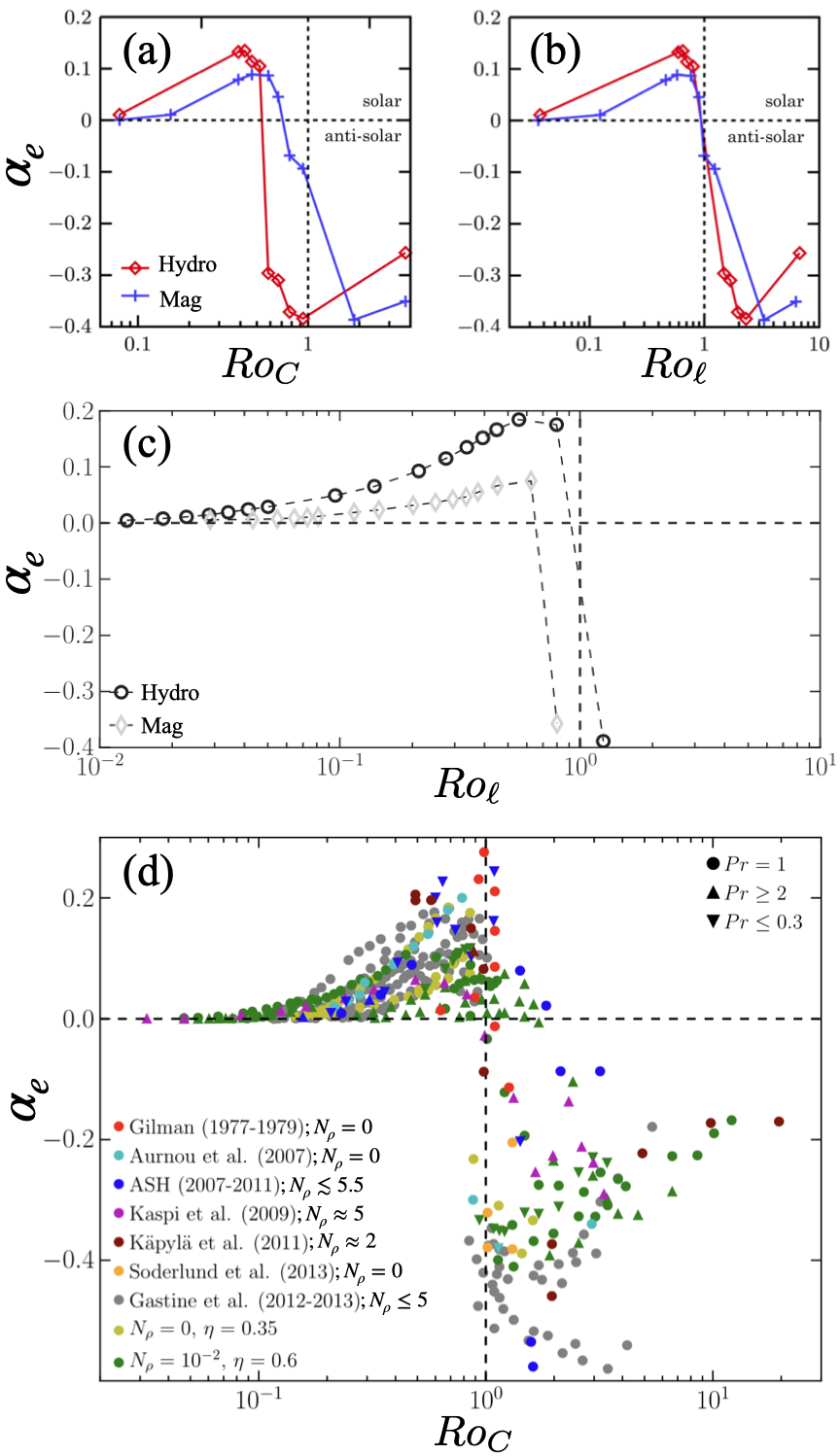}
\caption{Comparison of differential rotation. {\bf(a)} Equatorial surface zonal velocity, $\alpha_e$, as a function of convective Rossby number, $Ro_C$, in hydrodynamic (red) and magnetohydrodynamic dynamo (blue) simulations with approximately three density scale heights ($N_{\rho} \sim 3$), adapted from \citet{MabuchiEA15}. {\bf(b)} Identical to {\bf(a)}, but with local Rossby number, $Ro_\ell$, as the abscissa. {\bf(c)} Similar to {\bf(b)}, but showing Boussinesq ($N_{\rho}=0$) simulation results adapted from \citet{GastineEA14_MNRAS}. {\bf(d)} Compilation of hydrodynamic cases with background density stratifications ranging from $N_{\rho} = 0$ to 5.5 and with symbol shapes denoting the value of the Prandtl number, $Pr$, adapted from \citet{GastineEA14_MNRAS}.}
\label{fig:MabGast}
\end{figure}

\begin{figure*}
\noindent\includegraphics[width =36pc]{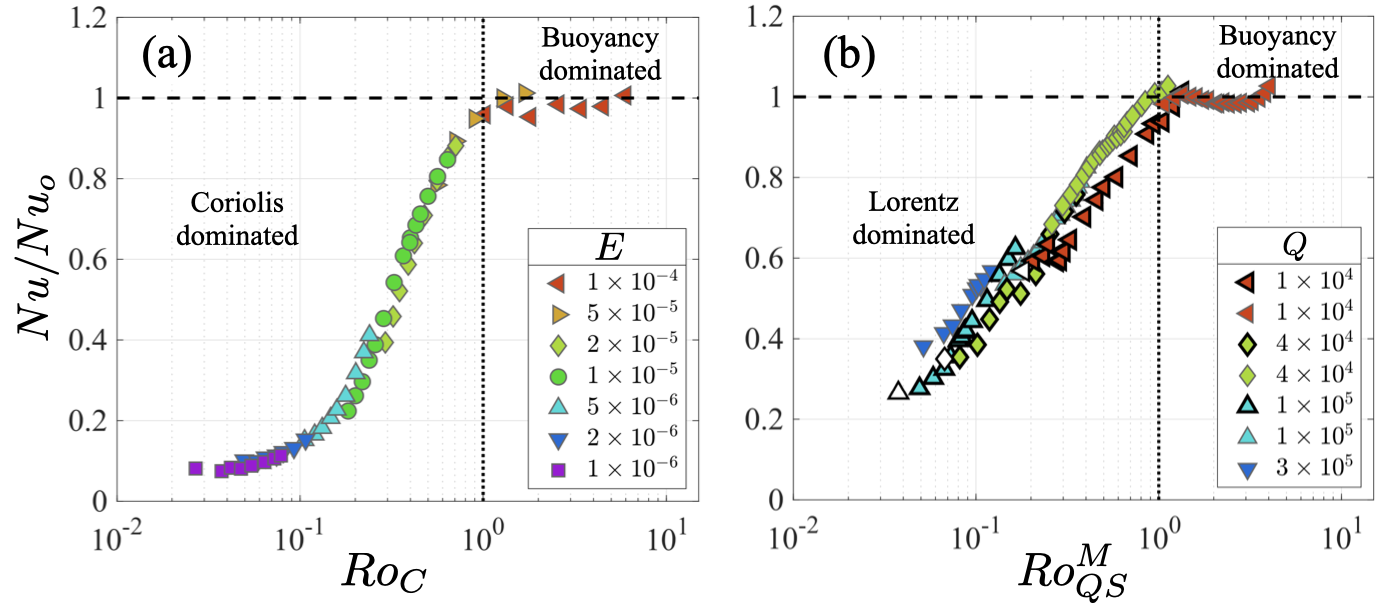}
\caption{Comparison of normalized heat transfer, $Nu/Nu_o$, adapted from \citet{KingAurnou15} and \citet{Xu23}. {\bf(a)} Rotating (non-magnetic) convective heat transfer as a function of convective Rossby number, $Ro_C$. Ekman number values are defined here as $E = \nu/(2 \Omega D^2)$. {\bf(b)} Magnetoconvective (non-rotating) heat transfer as a function of the quasi-static magnetic Rossby number, $Ro^M_{QS}$, defined in equation (\ref{eq:QSARoM}). Chandrasekhar numbers are denoted by $Q$.}
\label{fig:Xu23}
\end{figure*}

\begin{figure*}
\noindent\includegraphics[width =36pc]{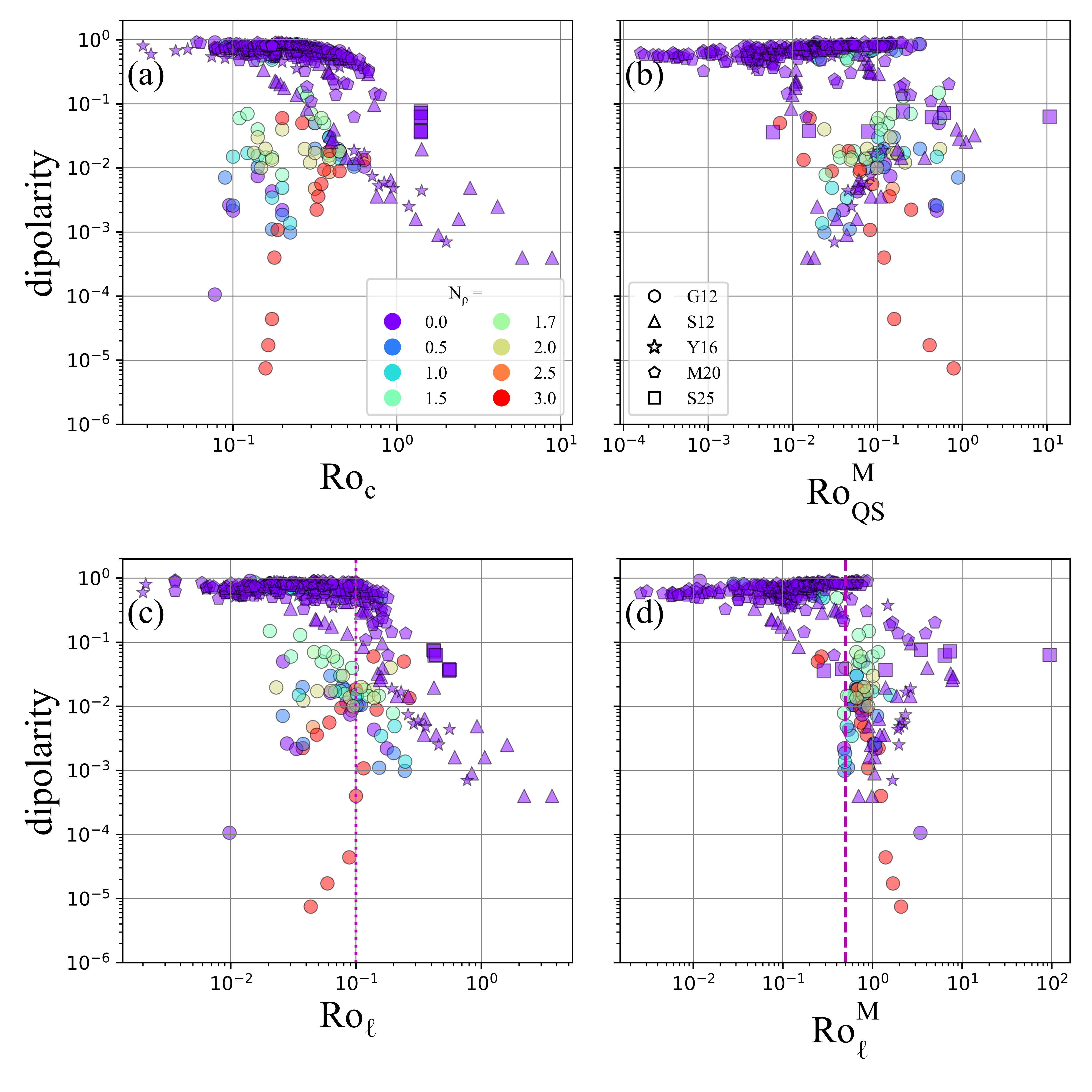}
\caption{Dipolarity, $f_D$, with data from \citet{GastineEA12} (G12); \citet{SoderlundEA12} (S12); \citet{Yadav_16} (Y16); \citet{Menu_20} (M20); and this study (S25), as a function of {\bf(a)} convective Rossby number, $Ro_C$, {\bf(b)} quasi-static magnetic Rossby number, $Ro^M_{QS}$, {\bf(c)} local Rossby number, $Ro_\ell$, and {\bf(d)} local magnetic Rossby number, $Ro_\ell^M$. Color refers to density stratification, $N_\rho$. Definitions for $f_D$, $Ro_\ell$, and $\Lambda_\ell$ used to calculate $Ro_\ell^M$ differ between studies, as described in Table~\ref{tab:definitions}; model set-ups also differ between these studies (e.g., both stress-free and no-slip models are shown).} 
\label{fig:G12}
\end{figure*}
 
In the last ten years, it has been posited in a number of studies that stellar differential rotation is primarily controlled by rotational hydrodynamics in the convection zone \cite[e.g.,][]{GastineEA13, GastineEA14_MNRAS, GuerreroEA13, KapylaEA14, MabuchiEA15, VivianiEA18, Camisassa22}. This hydrodynamic control is parameterized via the so-called convective Rossby number, $Ro_C$, which estimates the import of buoyant inertia relative to rotational inertia of a given system \citep[e.g.,][]{Gilman77, GilmanFoukal79, BrunPalacios09, SoderlundEA14, Soderlund19, AurnouEA20, VasilEA21}. 

Figure~\ref{fig:MabGast} shows normalized equatorial surface zonal velocities, 
\begin{equation}
\alpha_e = \overline{U}_\phi/(\Omega_o r_o), 
\label{eqn:alpha}
\end{equation}
in stellar-like convection zone simulations from (a,b) \citet{MabuchiEA15} and (c,d) \citet{GastineEA14_MNRAS}. Here, $\overline{U}_\phi$ is the mean azimuthal velocity of the surface at the equator, $\Omega_o$ is the mean angular rotation velocity, and $r_o$ is the outer boundary radius. (Note that $\alpha_e$ is synonymous with the equatorial Rossby number, $Ro_{e}$, often used in the geophysics literature.) These studies, amongst others, show that there is a transition in $\alpha_e$ between solar-like with a prograde equatorial jet and anti-solar with a retrograde equatorial jet that occurs near a Rossby number of order unity.

In Figure~\ref{fig:MabGast}a, the convective Rossby number, $Ro_C$, is plotted as the control parameter on the abscissa for hydrodynamic cases in red and magnetohydrodynamic (MHD) dynamo cases in blue. For $Ro_C \gtrsim 1$, this parameter estimates the ratio of the rotation time scale and the buoyant free-fall time across the fluid layer, $Ro_C \simeq U_{ff}/(\Omega D)$, where $U_{ff} \simeq \sqrt{(\beta \Delta T) g_o D}$ is the free-fall velocity \citep[][]{AurnouEA20}. The convective Rossby number can additionally be written in terms of typical dynamo control parameters: 
\begin{equation}
    Ro_C = \sqrt{\frac{Ra E^2}{Pr}}, \quad \text{where } Ra=\frac{\beta g_o \Delta T D^3}{\nu \kappa}, \quad E=\frac{\nu}{\Omega D^2}, \quad Pr= \frac{\nu}{\kappa}. 
    \label{eqn:RaEPr}
\end{equation} 
The Rayleigh number $Ra$ is the non-dimensional buoyancy forcing, the Ekman number $E$ is the non-dimensional rotation period, and the Prandtl number $Pr$ defines the fluid's thermomechanical properties. Here, $\beta$ is thermal expansivity, $g_o$ is gravitational acceleration at the outer boundary, $\Delta T$ is the superadiabatic temperature difference across the shell, $D = r_o - r_i$ is shell thickness where $r_i$ denotes the inner shell radius, $\nu$ is kinematic viscosity, and $\kappa$ is thermal diffusivity.

Figure~\ref{fig:MabGast}b shows the same $\alpha_e$ data plotted versus the \textit{a posteriori} local Rossby number, defined as 
\begin{equation}
Ro_\ell = \frac{U}{\Omega_o \ell_U}, 
\label{eqn:localRo}
\end{equation}
where $U$ is the characteristic velocity estimate and $\ell_U$ is the characteristic length scale of the velocity field \citep[e.g.,][]{ChristensenAubert06,Featherstone16, GuervillyEA19, AurnouEA20, OliverEA23}. In \cite{MabuchiEA15}, $U$ is the root-mean-square (rms) convective velocity that does not include the azimuthal flow components and they define $\ell_U$ based on the cross-axial scale of the largest identifiable convective eddies in each simulation. We also note that \citet{MabuchiEA15} has a factor of 1/2 in their definition of the Ekman number, which then reduces their values of $Ro_C$ and $Ro_\ell$ by 1/2 relative to the values reported in \citet{GastineEA14_MNRAS} and in this article.

Figure~\ref{fig:MabGast}c is adapted from \cite{GastineEA14_MNRAS} and shows $\alpha_e$ versus $Ro_\ell$ with hydrodynamic cases denoted with black circles and MHD dynamo cases with grey diamonds. In this study, $U$ is taken to be the rms equatorial azimuthal velocity and $\ell_U$ is the kinetic-energy-weighted length scale \citep{Christensen06}.
The transition between solar-like and anti-solar differential rotation in panels (a - c) occurs near to a hydrodynamic Rossby number value of order unity.

The different $U$ measurements in \cite{MabuchiEA15} and \cite{GastineEA14_MNRAS} do not qualitatively change the differential rotation transition points in panels (a - c). We postulate that this insensitivity to $U$ occurs because the different components of the velocity fluctuations in rapidly-rotating convection are all of the same order of magnitude \citep[e.g.,][]{StellmachEA14, Hadjerci24}. The same holds true in non-rotating turbulent convection, in which the bulk flow is nearly isotropic \cite[e.g.,][]{Nath16}.

Figure~\ref{fig:MabGast}d shows hydrodynamic spherical shell convection cases that span a wide range of radial background density stratifications from \cite{GastineEA14_MNRAS}. This stratification is quantified through $N_{\rho} = \ln (\rho_i / \rho_o)$, which measures the number of density scale heights across the fluid shell. Here, $\rho_i$ ($\rho_o$) is the background density at the inner (outer) shell boundary. The $N_\rho$ values range from 0 for Boussinesq cases up to approximately 5.5 in anelastic cases. Importantly, the transition between solar-like differential rotation at $Ro_C \lesssim 1$ to anti-solar differential rotation at $Ro_C \gtrsim 1$ is not strongly affected over this range of $N_\rho$. This implies that density stratification does not invalidate the $Ro_C \sim 1$ differential rotation regime boundary. We postulate that this occurs because the Rossby number, and hence $Ro_C$, does not directly depend on the mean fluid density. Instead, it is the characteristic velocity that tends to depend on density \citep{GastineEA13}. Thus, the effects of anelastic density stratification are accounted for implicitly in the Rossby number since it is proportional to the characteristic velocity as shown, for example, in equation (\ref{eqn:localRo}).

The modeling results shown in Figures~\ref{fig:MabGast}(a-c) have led to the idea that magnetohydrodynamic effects are subdominant to hydrodynamic processes in large-scale convection zone dynamics. However, the recent studies of \cite{Menu_20}, \cite{ZaireEA22}, \cite{hotta2022generation} and \cite{kapyla2023transition} all find that MHD effects cannot be neglected in convection zone dynamics. Thus, we hypothesize that the idea of purely hydrodynamic control of convection zone transitions may be a byproduct of the tendency to fix the value of electrical conductivity, $\sigma$, in these simulations. The nondimensional form of $\sigma$ is expressed via the magnetic Prandtl number: 
\begin{equation}
    Pm = \frac{\nu}{\eta} = \nu \mu_o \sigma,
    \label{eqn:Pm}
\end{equation} 
where $\eta = 1/(\mu_o \sigma)$ is the magnetic diffusivity and $\mu_o$ is the magnetic permeability. The magnetic Prandtl number is often kept close to unity ($Pm \simeq 1$) in stellar and planetary dynamo modeling efforts. This characteristic fixity is a consequence of dynamo simulations being the least numerically taxing to compute in the $Pm \simeq 1$ regime. In addition, turbulent diffusivity arguments often favor setting all diffusivities to comparable values such that the turbulent diffusivity ratios are of order unity \citep{HottaEA12, Roberts12}. This turbulent diffusivity argument must, however, be treated with care: the molecular diffusivities can take on a broad range of values across stellar convection zones and, thus, it is not obvious that the effective Prandtl numbers are everywhere of order unity \cite[cf.][]{Garaud21,PandeyEA2022,KapSin2022}. By fixing $\sigma$, or alternatively $Pm$, the strength of the Lorentz force will tend to remain approximately fixed in a given set of simulations. This could make it seem that MHD effects are not relevant for a given transition in the behavior of the system. 

We have focused on transitions in zonal flow behavior thus far. However, transitions in convective heat transfer have also been used to denote changes in global-scale behavioral dynamics \cite[e.g.,][]{plumley2019, Xu23}. This assumes that the mean heat transfer is a byproduct of all the collective convective motions that occur in a given fluid layer. Thus, changes in heat transfer scaling behaviors are argued to be related to changes in large-scale dynamics. In laboratory experiments where velocity field data has not been acquired \citep[e.g.,][]{KingAurnou13,Xu23}, it is necessary to make use of this heat transfer ansatz in order to cross-compare the convection regimes in these systems.

Figure~\ref{fig:Xu23} shows laboratory convective heat transfer measurements in a liquid gallium-filled cylinder heated from below and cooled from above as a function of (a) applied rotation rate and (b) applied magnetic field strength, adapted from \citet[][]{KingAurnou15} and \citet[][]{Xu23}. The convective heat transfer data, $Nu$, is normalized in both figure panels by the heat transfer measured in non-rotating, non-magnetic convection experiments, $Nu_o$. The rotating convection data in Figure~\ref{fig:Xu23}a shows that the convective heat transfer is reduced when $Ro_C \lesssim 1$. In this regime, rotation dominates over buoyancy-driven inertia and the flow becomes strongly rotationally-constrained \citep[e.g.,][]{Julien07}. It is found that $Nu/Nu_o \approx 1$ when $Ro_C \gtrsim 1$, showing that the convective flow is weakly constrained by rotation when the buoyancy effects exceed the Coriolis forces. 

The quasi-static ($QS$), low magnetic Reynolds number ($Rm = UD/\eta \ll 1$), liquid metal magnetoconvection data shown in Figure~\ref{fig:Xu23}b is similar in gross structure to that of Figure~\ref{fig:Xu23}a. The abscissa in Figure~\ref{fig:Xu23}b shows the magnetic Rossby number, which is the analog to $Ro_C$ for these conditions:
\begin{equation}
Ro^M_{QS} = \sqrt{\frac{Ra Q^{-2}}{Pr}}, \quad \text{where }  Q = \frac{\sigma B_o^2 D^2}{ \rho \nu}. 
\label{eq:QSARoM}
\end{equation} 
The Chandrasekhar number $Q$ estimates the ratio of Lorentz and viscous forces, with $B_o$ denoting the \textit{externally imposed} magnetic field strength and $\rho$ denoting the fluid density. 
This form of the magnetic Rossby number characterizes the ratio of inertial and Lorentz forces in the low $Rm$, quasi-static limit in which the current density is estimated via Ohm's Law, 
$\textbf{J} = \sigma \textbf{u} \times \textbf{B}$
\citep[e.g.,][]{sarris2006, yan2019, Xu23, horn2024}. 

In rotating systems, the quasi-static magnetic Rossby number can be recast as 
\begin{equation}
Ro^M_{QS}= \frac{Ro_C}{\Lambda_i}, \quad \mbox{where} \quad \Lambda_i = \frac{\sigma B^2}{\rho \Omega}.
\label{eq:Elsasser}
\end{equation}
Here, $\Lambda_i$ is the traditional form of the Elsasser number that estimates the ratio of Lorentz and Coriolis forces also in the low-$Rm$, quasi-static limit \citep{CardinEA02, SoderlundEA15, dormy2016, Aurnou17, horn2024}.

The normalized heat transfer in Figure~\ref{fig:Xu23}b is decreased in the $Ro^M_{QS} \lesssim 1$ regime, which shows that the quasi-static, low-$Rm$ Lorentz force constrains the convective turbulence when they exceed the strength of the inertial forces \citep[e.g.,][]{Julien07}. The similarities in zeroth-order structure of the data in Figures~\ref{fig:Xu23}a,b imply that Coriolis forces and Lorentz forces are both capable of changing the global-scale convection dynamics, which are parameterized in this figure by the global heat transferred through the fluid layer \citep[cf.][]{GlazierEA99}. 

\section{Local Magnetic Rossby Number}

We hypothesize that grossly similar behavioral transitions as in Figure \ref{fig:Xu23} also exist in high-$Rm$ spherical dynamo systems. Thus, we postulate that convection zone dynamics in spherical shell dynamo models will undergo regime transitions as a function of both the Rossby number and the high-$Rm$ form of the magnetic Rossby number. As discussed above, it is well known that $Ro_C$ correlates with changes in convection zone dynamics. We will therefore focus in this study on determining the conditions under which changes in Lorentz forces alter the convection zone dynamical regime. 

The magnetic Rossby number $Ro_{QS}^M$ defined in equation (\ref{eq:QSARoM}) is, however, not a well-defined control parameter in dynamo simulations. It cannot be calculated \textit{a priori} since the magnetic field strength is not known before running a given dynamo simulation. This differs from the laboratory magnetoconvection experiments shown in Figure~\ref{fig:Xu23}b in which the magnetic field is applied by an external magnet. 

\begin{table*}
\setlength{\tabcolsep}{3pt}
\renewcommand{\arraystretch}{1.2}
\centering 
\begin{tabular}{cc|ccccccccccccccccccc}
\hline 
Regime & $Pm$ & $Ro_\ell^M$ & $Rm$ & $Re$ & $Re_{m>0}$ & $Ro$ & $Ro_{l}$ & $\alpha_e$ & $\ell_U$ & $\ell_B$ & $\ell_U / \ell_B$ & $Nu$ & $q^{eqtr}_{rel}$ & $q^{pole}_{rel}$ & $f_D^{10}$ & $f_Q$ & $f_{A}$ & $\Lambda_i$ & $\Lambda_l$ & $\mathcal{E}_M$ / $\mathcal{E}_K$ \\ 
\hline  
SM-S    & 10 & 0.287    & 4348 & 435 & 418 & 0.130 & 0.551 & +0.041 & 0.237 & 0.029 & 8.25 & 12.8 & -0.26 & 1.00 & 0.036 & 0.058 & 0.508 & 240 & 1.92 & 0.42  \\   
SM-S    & 5  & 0.459    & 2263 & 453 & 417 & 0.136 & 0.552 & +0.062 & 0.246 & 0.033 & 7.45 & 12.7 & -0.28 & 1.17 & 0.038 & 0.055 & 0.503 &  90 & 1.20 & 0.29  \\   
SM-S    & 2  & 1.40     & 1068 & 534 & 457 & 0.160 & 0.561 & +0.076 & 0.286 & 0.042 & 6.86 & 12.6 & -0.20 & 0.77 & 0.037 & 0.058 & 0.493 &  18 & 0.400 & 0.10  \\  
\hline
EQ-AS   & 1    & 3.45  & 986  & 986  & 754 & 0.296 & 0.416 & -0.287 & 0.711 & 0.058 & 12.2 & 13.5 & 0.66 & 0.40 & 0.076 & 0.11 & 0.662 & 7.0 & 0.121 & 0.024   \\   
EQ-AS   & 0.75 & 6.45  & 756  & 1008 & 735 & 0.302 & 0.423 & -0.303 & 0.714 & 0.064 & 11.1 & 13.3 & 0.57 & 0.68 & 0.062 & 0.14 & 0.800 & 3.2 & 0.066 & 0.014   \\   
\hline
AQ-AS    & 0.50  & 7.31 & 499  & 998  & 697 & 0.299  & 0.419 & -0.313 & 0.714 & 0.082 & 8.76 & 13.1 & 0.41 & 0.81 & 0.072 & 0.20 & 1.04 & 2.3 & 0.057 & 0.016   \\   
AQ-AS    & 0.25  & 93.7 & 269  & 1077 & 692 & 0.323  & 0.435 & -0.341 & 0.743 & 0.104 & 7.14 & 13.1 & 0.26 & 1.26 & 0.063 & 0.27 & 1.27 & 0.13 & 0.005 & 0.002  \\   
\hline
\end{tabular}
\caption{Diagnostic parameters for variable $Pm$ cases all with fixed $\chi=0.35$ and $Ro_C = 1.4$ (corresponding to $Ra=2.22 \times 10^{7}$, $E=3.0 \times 10^{-4}$, and $Pr=1$). Parameters are defined in the text. Dynamo models with $Pm \geq 2$ are in the strongly multipolar, solar-like differential rotation (SM-S) regime; models with $Pm = [0.75, 1]$ are in the equatorial quadrupole, anti-solar (EQ-AS) regime; and models with $Pm \leq 0.5$ are in the axial quadrupole, anti-solar (AQ-AS) regime.}
\label{tab:Pmruns}
\end{table*}
    
Furthermore, since $Rm > \mathcal{O}(10)$ in dynamo systems \citep{RobertsKing13}, $Ro^M_{QS}$ cannot be employed to interpret dynamo modeling results. Instead, we will use a generalized form of the magnetic Rossby number that holds in $Rm > 1$ dynamo settings. Following \cite{SoderlundEA15}, the current density represented as $\textbf{J} = (\nabla \times \textbf{B})/\mu_o$ is valid for all $Rm$ values so long as the displacement currents are negligible \citep{Davidson01}. Further, velocity and magnetic field gradient operators are scaled in terms of $\ell_U^{-1}$ and $\ell_B^{-1}$, respectively (see Table~\ref{tab:definitions} for definitions used in the literature). This local-scale magnetic Rossby number then takes the form 
\begin{equation}
    Ro_\ell^M =   \frac{|\rho \textbf{u} \cdot \nabla \textbf{u}|} {|[(\nabla \times   \textbf{B})/\mu_o] \times \textbf{B}|} 
    \ \approx  \ \frac{U^2 / \ell_U}{[B^2 / (\rho \mu_o)] /\ell_B}
    \ \approx  \ \left(\frac{U^2}{V_A^2} \right) \left(\frac{\ell_B}{\ell_U} \right) 
    \label{eq:RoM1}
\end{equation}
where $V_A = \sqrt{B^2 / (\rho \mu_o)}$ is the Alfv\'en velocity \citep{schaeffer2012}.

Based on the rightmost expression in equation (\ref{eq:RoM1}), the local magnetic Rossby number is seen to be the kinetic to magnetic energy density ratio $(U^2 / V_A^2)$ modified by the ratio of dynamical length scales $(\ell_B / \ell_U)$. Thus, if $\ell_B \approx \ell_U$, then the ratio of the inertial and Lorentz forces is well approximated by the ratio of the kinetic and magnetic energy densities. In cases where $\ell_B$ and $\ell_U$ are not comparable, as likely occurs in dynamo simulations in which $Pm$ is not close to unity, then the inclusion of the $(\ell_B / \ell_U)$ term in $Ro_\ell^M$ is obligatory. 

Accurate $a \ priori$ estimates of $B$, $U$, $\ell_U$, and $\ell_B$ are difficult to make in turbulent dynamo systems. Thus, we recast $Ro_\ell^M$ in terms of $a \ posteriori$ output parameters $Ro_\ell$ (see equation~(\ref{eqn:localRo})) and the local Elsasser number $\Lambda_\ell$ in order to obtain a generalized (non-quasi-static, non-low $Rm$) local magnetic Rossby number:
\begin{equation}
    Ro_\ell^M = \frac{Ro_\ell}{\Lambda_\ell}, \quad \text{ where } \quad \Lambda_\ell = V_A^2 / (U \Omega \ell_B). 
    \label{eq:RoMell}
\end{equation}
Note that $Ro_\ell^M$ is not a true control parameter, unlike $Ro_{QS}^M$, since $Ro_\ell^M$ can only be calculated after a given dynamo simulation run has completed. Instead, it is better to think of $Ro_\ell^M$ as a diagnostic parameter. 

Figure \ref{fig:G12} provides an example of the utility of $Ro_\ell^M$ using data harvested from published datasets \citep{GastineEA12,SoderlundEA12,Yadav_16,Menu_20} and this study. The ordinate in all four panels denotes the values of dipolarity, $f_D$, defined in \citet{GastineEA12} as the ratio of the dipole to total magnetic energy on the model's outer spherical boundary (see Table~\ref{tab:definitions} for other definitions used in these studies). Here, dipolar magnetic field morphologies are defined to have $f_D \gtrsim 0.1$ and multipolar morphologies to have $f_D \lesssim 0.1$.

Bistable solutions exist in \citet{GastineEA12}: for the same control parameters, both dipolar and multipolar stable solutions are found to exist. Figures~\ref{fig:G12}a,b show that the rotational and magnetic control parameters $Ro_C$ and $Ro_{QS}^M$, respectively, are non-elucidatory here since the dipolarity values are rather strongly scattered in both panels. Further, the \textit{a posteriori} local diagnostic parameter $Ro_\ell$ also is unable to reduce the spread of $f_D$ data in Figure~\ref{fig:G12}c, despite $Ro_\ell \sim 0.1$ often serving as a proxy for predicting dipolar versus multipolar dynamos. In contrast, the local magnetic Rossby number $Ro_\ell^M$, plotted along the abscissa in Figure~\ref{fig:G12}d, adequately collapses the $f_D$ data, with dipolar solutions corresponding to strong local-scale Lorentz forces and with multipolar solutions clustered in the vicinity of $Ro_\ell^M \gtrsim 0.5$ such that their inertial and magnetic forces are approximately in balance. Thus, Figure~\ref{fig:G12} demonstrates the diagnostic capabilities and potential relevance of $Ro_\ell^M$ in analyzing the physics of dynamo systems.

Because one cannot calculate $Ro_\ell^M$ \textit{a priori}, we hold all control parameters fixed except for the magnetic Prandtl number, $Pm \propto \sigma$, in our dynamo simulations. This allows us to study how the relative strength of the Lorentz force correlates with the dynamo morphology, differential rotation, and surface heat flux patterns in our models \cite[e.g.,][]{Fan14, kapyla2017prandtl, AugustsonEA19, Menu_20, ZaireEA22}. We find distinct behavioral states as a function of $Ro_\ell^M$. This implies that not only is the convective Rossby number important to the large-scale convection zone dynamics, but that its MHD counterpart, the local magnetic Rossby number $Ro_\ell^M$, is of dynamical importance in dynamo systems as well.

\begin{table*}
\setlength{\tabcolsep}{3pt}
\renewcommand{\arraystretch}{1.2}
\centering 
\begin{tabular}{l|llllll}
\hline 
Regime & Magnetic field morphology & Differential rotation & Heat transfer & $Ro_\ell^M$ ($Pm$) domain & Interpretation \\ 
\hline  
SM-S  & Strongly-Multipolar (SM)                                & Solar-like (S)     & Polar cooling                          & $Ro_\ell^M \lesssim 1.4$ & Lorentz dominated \\
      & $f_Q \sim 0.06$, $f_{A} \sim 0.5$, $f_D^{10} \sim 0.04$ & $\alpha_e > 0$     & $q_{eqtr} < q_{mid-lat} < q_{pole}$    & ($Pm \geq 2$) &  \\
\hline  
EQ-AS & Equatorial Quadrupole (EQ)                              & Anti-Solar (AS)    & Polar \& Equatorial cooling            & $3.5 \lesssim Ro_\ell^M \lesssim 6.4$ \\
      & $f_Q \sim 0.1$, $f_{A} \sim 0.7$, $f_D^{10} \sim 0.07$  & $\alpha_e < 0$     & $q_{mid-lat} < q_{eqtr} \sim q_{pole}$ & ($0.75 \leq Pm \leq 1$) & \\
\hline  
AQ-AS & Axial Quadrupole (AQ)                                   & Anti-Solar (AS)    & Polar (\& Equatorial) cooling          & $Ro_\ell^M \gtrsim 7.3$ & Buoyancy dominated\\   
      & $f_Q \sim 0.2$, $f_{A} \sim 1$, $f_D^{10} \sim 0.07$  & $\alpha_e < 0$     & $q_{mid-lat} < q_{eqtr} < q_{pole}$      & ($Pm \leq 0.5$) & \\   
\hline
\end{tabular}
\caption{Definition of regimes identified in our study with fixed $Ro_C = 1.4$ based on the magnetic field, differential rotation, and heat transfer characteristics.}
\label{tab:regimes}
\end{table*}

\begin{figure*}
\includegraphics[width=\textwidth]{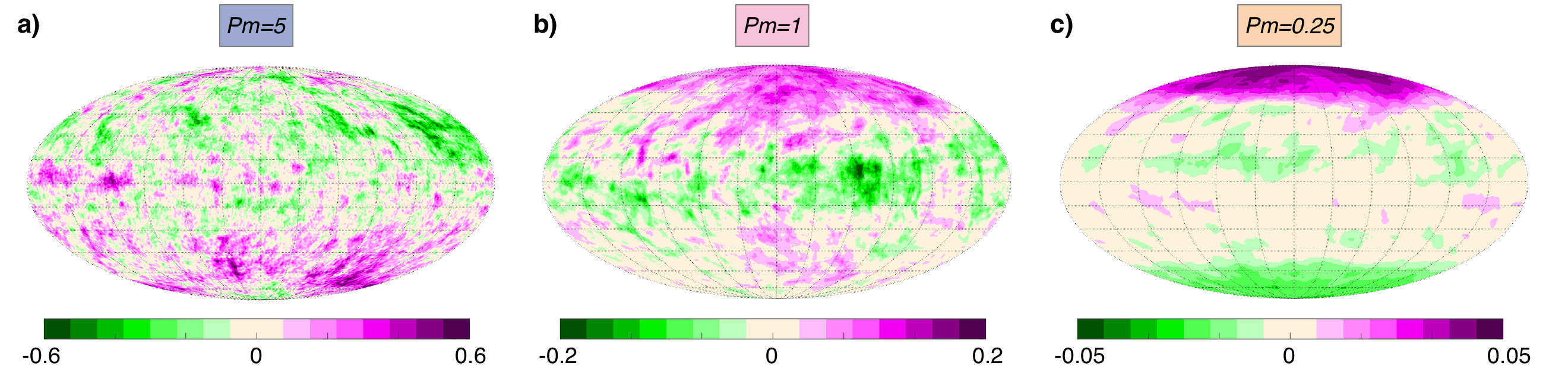}
\caption{Time-averaged radial magnetic fields on the outer boundary for the {\bf(a)} $Pm=5$, {\bf(b)} $Pm=1$, and {\bf(c)} $Pm=0.25$ models in units of $\sqrt{\Lambda_i}$. Purple denotes outward directed fields, while green denotes inward directed fields.} 
\label{fig:SurfBr}
\end{figure*}

\section{Methods}

We use the open-source, pseudospectral dynamo code MagIC \citep{Wicht02,GastineWicht12} with the SHTns library to efficiently calculate the spherical harmonic transforms \citep{Schaeffer13}. The models carried out here simulate three-dimensional (3D), time-dependent, thermally-driven convection of a Boussinesq fluid in a spherical shell rotating with constant angular velocity $\Omega {\hat{\bf z}}$. The shell's geometry is set to $\chi = r_i / r_o = 0.35$ with boundaries that are isothermal, impenetrable, and stress-free. The inner sphere, $r < r_i$, is treated as a solid that has the same electrical conductivity as the fluid shell, the outer boundary of which is electrically insulating. Gravity varies linearly with spherical radius. The dimensionless governing equations for this system are 
\begin{equation}
 E \left(\frac{\partial \textbf{u}}{\partial t} + \textbf{u} \cdot
\nabla \textbf{u}-\nabla^2 \textbf{u}\right)+ 2\hat{{\bf z} } \times
\textbf{u} + \nabla p 
    =  \frac{Ra E}{Pr} \frac{{\bf r}}{r_o}T +
    \frac{1}{Pm} (\nabla \times \textbf{B}) \times \textbf{B} ,
    \label{eqn:NS}
    \end{equation}
\begin{equation}
\frac{\partial \textbf{B}}{\partial t} = \nabla \times (\textbf{u}
\times \textbf{B}) + \frac{1}{Pm} \nabla^2 \textbf{B},
    \label{eqn:ind}
\end{equation}
    \begin{equation}
\frac{\partial T}{\partial t} + \textbf{u} \cdot \nabla T =
    \frac{1}{Pr} \nabla^2 T, 
    \end{equation}
\begin{equation}
\nabla \cdot \textbf{u}  = 0, \indent{\nabla \cdot \textbf{B}}  = 0, 
\end{equation}
where $\textbf{u}$ is the velocity vector, $\textbf{B}$ is the magnetic induction vector, $T$ is the temperature, and $p$ is the non-hydrostatic pressure. We make use of typical non-dimensionalizations used in the dynamo literature: $D$ as length scale; $\Delta T$ as temperature scale; $\tau_{\nu} \sim D^{2}/ \nu$ as time scale; $\rho \nu \Omega$ as pressure scale; $\nu/D$ as velocity scale such that the non-dimensional rms flow speed is equal to the Reynolds number $Re=UD/\nu$; and $\sqrt{\rho \mu_{o} \eta \Omega}$ as magnetic induction scale such that the square of the non-dimensional rms magnetic field strength is equal to the traditionally-defined Elsasser number $\Lambda_{i}$. 

The governing non-dimensional parameters are the radius ratio $\chi$, the magnetic Prandtl number $Pm$, the thermal Prandtl number $Pr$, the Ekman number $E$, and the Rayleigh number $Ra$ (see Eqs.~\ref{eqn:RaEPr} and ~\ref{eqn:Pm}). For all models carried out herein, we use fixed values of $E=3.0 \times 10^{-4}$, $Ra=2.22 \times 10^7$, and $Pr=1$, yielding a fixed convective Rossby number of $Ro_C = 1.4$. For $Ro_C \gtrsim 1$, the buoyancy forces tend to overwhelm the Coriolis forces, generating quasi-3D convective turbulence which typically act to generate anti-solar differential rotation profiles \citep[e.g.,][]{AurnouEA07, SoderlundEA13, GastineEA13, GastineEA14_MNRAS, FeatherstoneMiesch15, MabuchiEA15, Soderlund19}. We have chosen to fix $Ro_C \approx 1$ in order to be close to the rotationally-controlled zonal flow transition point (see Figure~\ref{fig:MabGast}).

This choice of $Ro_C \approx 1$ should position our dynamo survey such that it is sensitive to any possible magnetically-controlled dynamical transitions. Towards this end, we systematically vary the electrical conductivity such that $Pm=[10, 5, 2, 1, 0.75, 0.50, 0.25]$, noting that dynamo action becomes subcritical for $Pm \lesssim 0.2$. As shown in Table~\ref{tab:Pmruns}, the resulting $Ro_\ell^M$ values range from 0.3 to 94 and thus cross unity, where a transition may be most intuitive. 

The choice of a Boussinesq fluid in a thick fluid shell with isothermal thermal boundary conditions defines as simple a system as could be created for this problem, both physically and computationally. We argue that this simplified approach is appropriate for the determination of these zeroth-order hydrodynamic regime transitions \citep[e.g.,][]{Gilman77, AurnouEA07}. In particular, hydrodynamic studies have been found, to date, to be only weakly sensitive to anelastic effects, shell geometry, and style of buoyancy forcing \citep[e.g.,][]{GastineEA13, GastineEA2014pepi, GastineEA14_MNRAS, FeatherstoneMiesch15, Gastine23, LemasquerierEA23}. In contrast, dynamo systems show some sensitivity to both MHD effects and anelasticity \citep[e.g.,][]{AugustsonEA19, hotta2022generation, ZaireEA22}, but this broad parameter space is not well characterized. 

The models use 192 spherical harmonic modes, 65 radial levels in the outer shell, and 17 radial levels in the inner core. No azimuthal symmetries or hyperdiffusivities are employed. All cases are initialized either using the results of prior dynamo models with different $Pm$ values or from random thermal perturbations and a seed magnetic field; the choice of initial conditions was found to have no significant effect on the results. This lack of hysteresis is consistent with other MHD studies \citep[e.g.,][]{KarakEA15}, in contrast to the bistability identified in hydrodynamic simulations \citep[e.g.,][]{GastineEA14_MNRAS,KapylaEA14}. Once the initial transient behavior has subsided, the model results are all time-averaged over a time window $\Delta t$. This corresponds to the non-dimensional averaging time $\Delta t_\nu^* = \frac{\Delta t} {\tau_\nu} = C = 0.09$ measured in viscous diffusion time units. Expressing this non-dimensional averaging window in magnetic diffusion times ($\tau_\eta = D^2/\eta$) and convective overturn times ($\tau_U = D/U$), respectively, yields 
\begin{equation}
        \Delta \tau_\eta^* = \frac{\Delta t} {\tau_\eta} = \frac{\Delta t} {\tau_\nu} \, \frac{\tau_\nu}{\tau_\eta} 
    = C \, \frac{\eta}{\nu} = 0.09 \, Pm^{-1} \, ,
\end{equation}
\begin{equation}
    \Delta t_U^* = \frac{\Delta t} {\tau_U} = \frac{\Delta t} {\tau_\nu} \, \frac{\tau_\nu}{\tau_U} 
    = C \, \frac{U D}{\nu}= 0.09 \, Re \, .
\end{equation}
Thus, $\Delta t_\eta^*$ ranges from 0.003 to 0.4 and the associated $\Delta t_U^*$ values range from 39 to 97 (see Table~\ref{tab:Pmruns}).

\begin{figure*}
\center{
\noindent\includegraphics[width = 36pc]{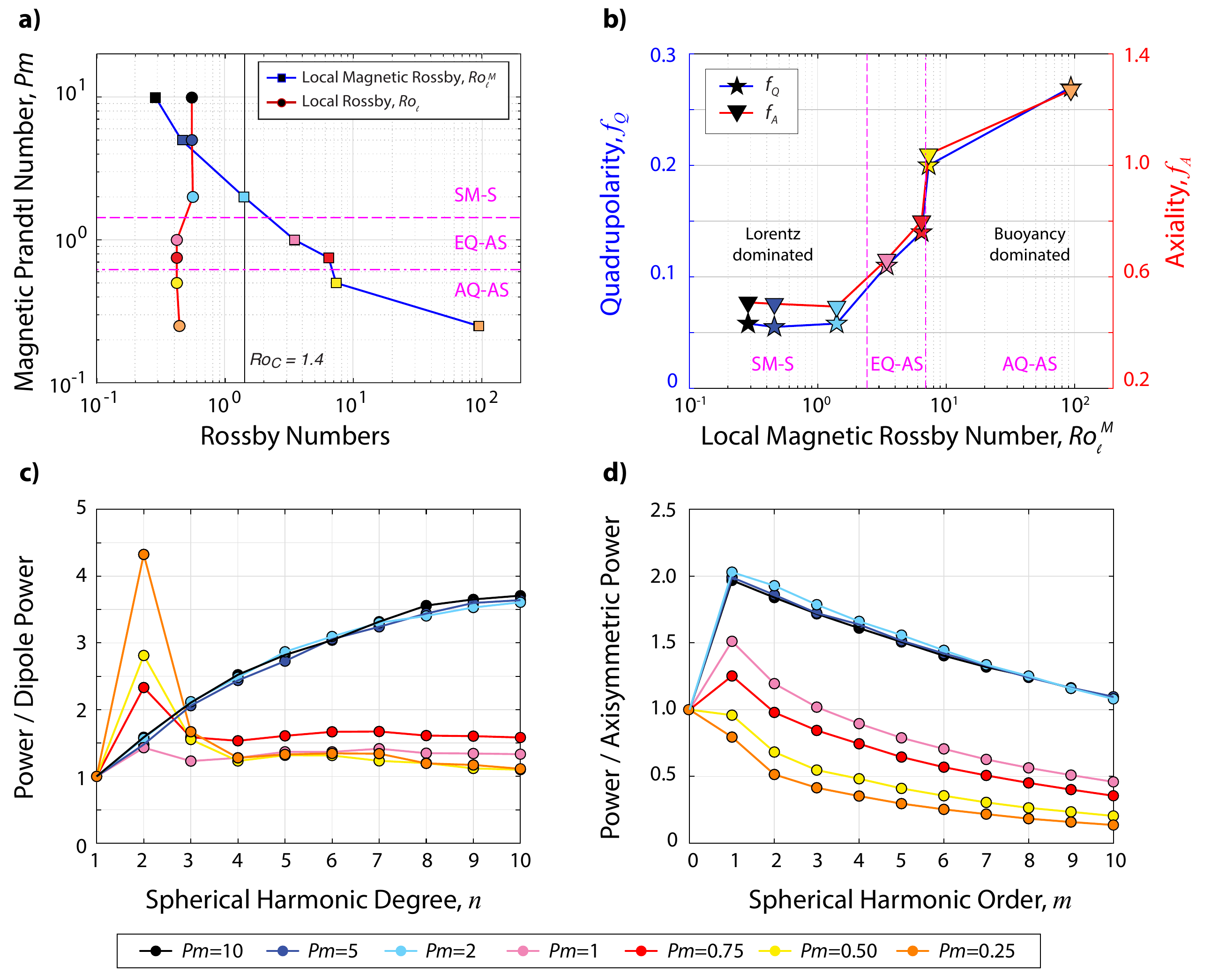} 
\caption{Time-averaged magnetic field characteristics. {\bf (a)} Local magnetic Rossby $Ro_\ell^M$ (blue line) and local hydrodynamic Rossby $Ro_{\ell}$ (red line) numbers versus the magnetic Prandtl number, $Pm$. For comparison, the fixed convective Rossby number of $Ro_C=1.4$ is denoted by the black line. {\bf (b)} Magnetic field morphology as measured by quadrupolarity $f_Q$ (blue line) and axiality $f_{A}$ (red line) versus the local magnetic Rossby number, $Ro_\ell^M$. Vertical magenta lines demarcate approximate regime boundaries. {\bf (c)} Magnetic power spectra as a function of spherical harmonic degree $n$, normalized by the $n=1$ dipole power. {\bf (d)} Magnetic power spectra as a function of spherical harmonic order $m$, normalized by the $m=0$ axisymmetric power. In all panels, color denotes $Pm$.} 
\label{fig:Bmorph}}
\end{figure*}

\section{Results}
Our analysis focuses on characteristics of the resulting magnetic fields (Section~\ref{sec:Mag}), velocity fields (Section~\ref{sec:Vel}), and convective heat transfer (Section~\ref{sec:HF}), with the results summarized in Tables~\ref{tab:Pmruns} and~\ref{tab:regimes}. As we will demonstrate, changing only the electrical conductivity of the fluid, as represented by $Pm$ ({\it a priori}) and $Ro_\ell^M$ ({\it a posteriori}), leads to first-order changes in the magnetic field, velocity field, and temperature field. Analysis of time-averaged magnetic spectra indicates three magnetic regimes: strongly-multipolar (SM), equatorial quadrupole (EQ), and axial quadrupole (AQ). Two types of differential rotations are found: solar-like (S) and anti-solar (AS). Latitudinal patterns of heat flux have three flavors: dominant polar heat flux and minimal equatorial heat flux, comparable polar and equatorial heat fluxes, or dominant polar heat flux with secondary peaks near the equator. We, therefore, define three dynamical regimes based on the dynamo and differential rotation: SM-S (strongly multipolar, solar-like), EQ-AS (equatorial quadrupole, anti-solar), and AQ-AS (axial quadrupole, anti-solar); see Table~\ref{tab:regimes}.

\begin{figure*}
\center{
\noindent\includegraphics[width =36pc]{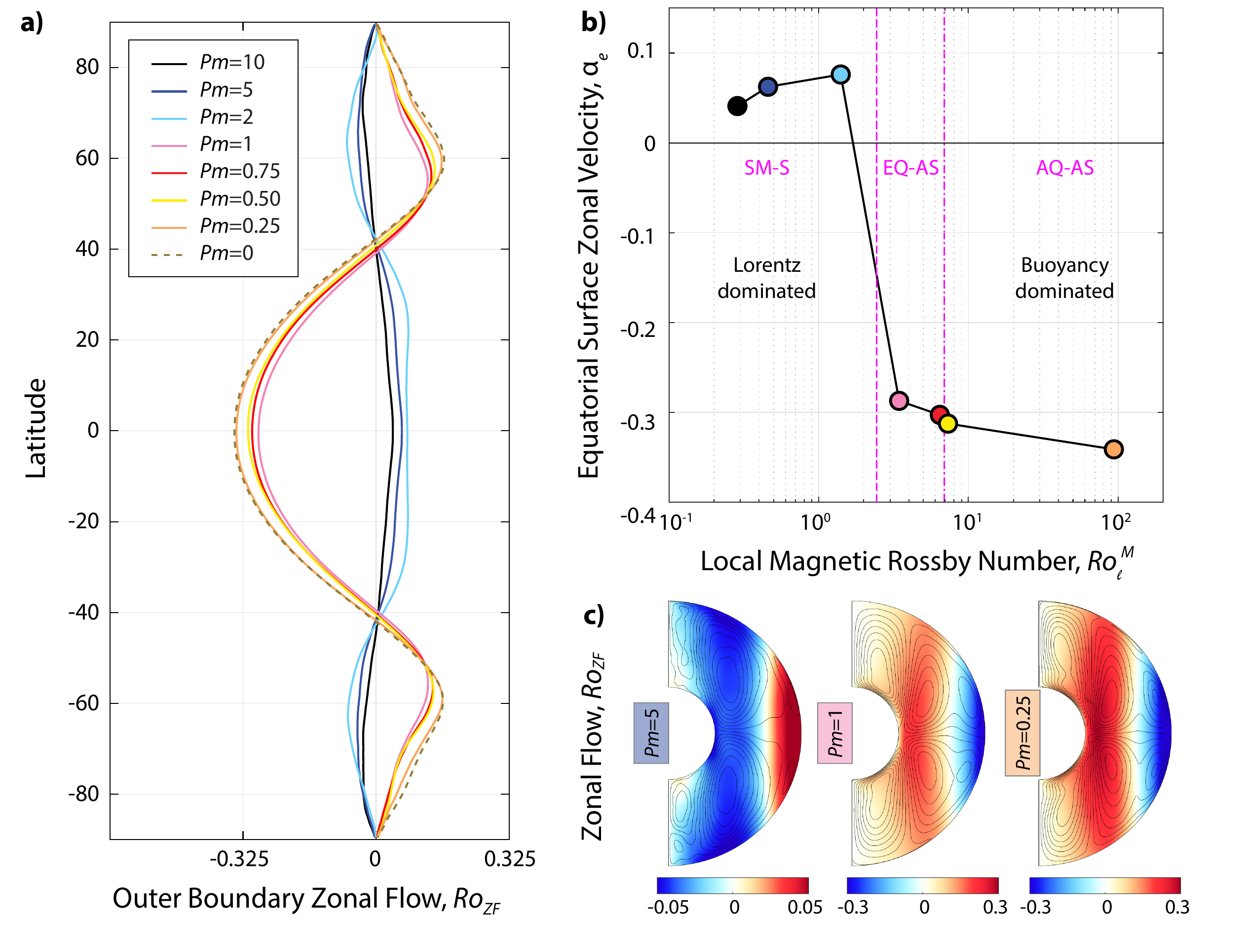}
\caption{Mean zonal flow characteristics, averaged over azimuth and time. {\bf (a)} Zonal flow profiles on the outer shell boundary as a function of latitude in Rossby number units, $Ro_{ZF}=\overline{U_{\phi}}/(\Omega r_o)$. Color denotes $Pm$. {\bf (b)} Equatorial zonal flow velocity ($\alpha_e$) from {\bf (a)} plotted versus the local magnetic Rossby number. Vertical magenta lines demarcate approximate regime boundaries. {\bf (c)} Zonal flows ($Ro_{ZF}$) and meridional circulation patterns for the (left) $Pm=5$, (middle) $Pm=1$, and (right) $Pm=0.25$ cases. Red (blue) indicates the prograde (retrograde) flow direction, while solid (dashed) black contours denote clockwise (counterclockwise) circulations.} 
 \label{fig:ZF_Pm}}
 \end{figure*}

\subsection{Magnetic Fields}
\label{sec:Mag}

The traditional Elsasser number increases from $\Lambda_i \sim 10^{-1}$ for $Pm=0.25$ to $\Lambda_i \sim 10^2$ for $Pm=10$ in a nearly monotonic trend (in log-log space). The local Elsasser number is approximately two orders of magnitude smaller, with $\Lambda_\ell \geq 0.4$ in the SM-S regime with $Pm \geq 2$ and $\Lambda_\ell \lesssim 0.1$ elsewhere. The characteristic length scale for the magnetic field is defined as 
\begin{equation}
    \ell_B = \pi/(2 \overline{k_B}), \quad \text{where } \quad \overline{k_B} = \sqrt{\overline{n_{B}}^2 + \overline{m_{B}}^2}
\end{equation}
with 
\begin{equation}\label{eq:magl}
    \overline{n_{B}} = {\sum_{\substack{n}}} \frac{n \langle {\bf B}_{n} \cdot {\bf B}_{n} \rangle}{\langle {\bf B} \cdot {\bf B} \rangle} \quad \text{and} \quad 
    \overline{m_{B}} = {\sum_{\substack{m}}} \frac{m \langle {\bf B}_{m} \cdot {\bf B}_{m} \rangle}{\langle {\bf B} \cdot {\bf B} \rangle}.
\end{equation}
following \citet{SoderlundEA12}. The smallest length scales are found in the cases with the strongest magnetic fields while the largest length scales are found in the magnetically weakest case by nearly a factor of four difference. For all models, the magnetic length scale is substantially smaller than the velocity length scale $\ell_U$ that will be discussed further in the next subsection, with $\ell_U / \ell_B$ being slightly larger than 10 for $Pm=[0.75,1]$ and slightly smaller than 10 otherwise. Similarly, the magnetic energy density is smaller than the kinetic energy density by at least a factor of two and up to nearly $10^3$ in the lowest $Pm$ case.

Recall that the local magnetic Rossby number can be expressed as the ratio of kinetic to magnetic energies times the ratio of magnetic to kinetic length scales as defined in equation~(\ref{eq:RoM1}). Given the change in both quantities (see Table~\ref{tab:Pmruns}), the local magnetic Rossby number ranges from $Ro_\ell^M = 0.3$ in the $Pm=10$ case to 94 in the $Pm=0.25$ case, which contrasts with the fixed $Ro_C =1.4$ convective Rossby number. These results suggest that a triple balance between buoyancy, Coriolis, and Lorentz must instead be considered. 

These dynamo differences are also illustrated by their radial magnetic fields, shown at the outer boundary in Figure~\ref{fig:SurfBr} for three representative electrical conductivity values. The $Pm=5$ case (panel a) has a relatively small-scale magnetic field with many positive and negative flux patches distributed across the surface. The intermediate $Pm=1$ case (panel b) has a notably larger-scale magnetic field with negative flux predominantly at low latitudes and positive flux more concentrated in the northern hemisphere. A sharper contrast is obtained for the lowest $Pm=0.25$ case (panel c) where magnetic intensity is strongest at high latitudes with much weaker field elsewhere. Thus, the radial magnetic field becomes larger scale as the magnetic Prandtl number decreases, or equivalently, as the local magnetic Reynolds number decreases. The larger-scale field, however, has lower intensities: the decrease in $Pm$ by a factor of 20 (which yields an $Ro_\ell^M$ increase by a factor of 8) from panels a to c leads to an approximate order of magnitude decrease in $|{\bf B}|$.

Magnetic field morphology is further quantified in Figure~\ref{fig:Bmorph} based on the behavior of the magnetic power spectra of the fluid shell up to spherical harmonic degree ($n$) and order ($m$) 10 shown in panels (c and d). The spectra demonstrate that models with $Pm \geq 2$, or $Ro_\ell^M \lesssim 1$, are characterized by strongly multipolar dynamos with significant power over a broad range of spherical harmonic degrees, as indicated by the increase in magnetic power from $n=1$ to $n=10$. For these models, the axisymmetric ($m=0$) contribution is distinctly small compared to the peak equatorially symmetric mode, and there is a gradual decay from $m=1$ towards smaller scales with higher $m$ values. Conversely, models with $Pm \leq 0.5$, or $Ro_\ell^M \gtrsim 7$, are characterized by a substantial quadrupole ($n=2$) component and have peak power in the axisymmetric $m=0$ mode. The intermediate $Pm=[0.75, 1.0]$ models ($3.5 \lesssim Ro_\ell^M \lesssim 6.4$) show combinations of these behaviors. The $Pm=1$ case (shown in pink) has a relatively flat spectra as a function of $n$ with the highest amplitude (marginally) at $n=2$ and a prominent peak at $m=1$, while the $Pm=0.75$ case (shown in red) has a more substantial peak at $n=2$ and a less substantial peak at $m=1$.

These behaviors are quantified through the dipolarity ($f_D^{10}$), quadrupolarity ($f_Q$), and axiality ($f_{A}$) parameters:
\begin{equation}
    f_D^{10} = \frac{\mathcal{P}(n=1)}{\Sigma_{n=1}^{n=10} \mathcal{P(\mbox{\it n})}}, \quad 
    f_Q = \frac{\mathcal{P}(n=2)}{\Sigma_{n=1}^{n=10} \mathcal{P(\mbox{\it n})}}, \quad
    f_{A} = \frac{\mathcal{P}(m=0)}{\mathcal{P}(m=1)},
\end{equation}
where $\mathcal{P}$ is magnetic power at the specified $n$ and $m$ values. Note that dipolarity here only includes magnetic power up to $n=10$ in the denominator (cf. Table~\ref{tab:definitions}) and remains low across the survey (Table~\ref{tab:Pmruns}). Three regimes are identified in Figure~\ref{fig:Bmorph}b, based on the slopes of quadrupolarity and axiality as a function of the local magnetic Rossby number. For $Ro_\ell^M \gtrsim 1$, the values of $f_Q$ and $f_{A}$ do not change substantially, and these cases are hereafter referred to as ``strongly multipolar (SM)". As $Ro_\ell^M$ increases beyond unity, the magnetic field becomes more quadrupolar and more axially aligned. Despite a small difference in $Ro_\ell^M$ between the $Pm=0.75$ and $Pm=0.50$ cases, a sharp jump occurs in both $f_Q$ and $f_{A}$ that distinguishes the ``equatorial quadrupole" (EQ) and ``axial quadrupole" (AQ) regimes.

\subsection{Velocity Fields}
\label{sec:Vel}

Table~\ref{tab:Pmruns} and Figures~\ref{fig:Bmorph}a and \ref{fig:ZF_Pm} show the characteristics of the velocity fields in our models. Despite the convective Rossby number being fixed at $Ro_C \sim 1$, globally-averaged flow speeds as measured by the Reynolds number, $Re$, differ by more than a factor of two across the survey. A less substantial increase in convective flow speeds is also identified when only the non-axisymmetric kinetic energy component is included, as measured by $Re_{m>0}$. The $Re_{m>0}/Re$ ratio changes between regimes. For cases with $Ro_\ell^M \lesssim 1$, corresponding to the strongly multipolar magnetic regime, $Re_{m>0}/Re > 0.85$ such that axisymmetric zonal flows and meridional circulations have a small contribution. This ratio decreases with increasing $Ro_\ell^M$ values, reaching $Re_{m>0}/Re = 0.64$ in the highest $Ro_\ell^M = 94$ case such that axisymmetric contributions become more significant. This contrast between global and convective Reynolds numbers indicates differences in the differential rotation and, secondarily, meridional circulations. 

In Figure \ref{fig:Bmorph}a we also show the local Rossby number, $Ro_\ell$, computed here assuming the \citet{GastineEA12} definition for typical flow length scale:
\begin{equation}\label{eq:vell}
    \ell_U = \pi / \overline{n_{U}}, \quad \text{where } \quad \overline{n_{U}} = {\sum_{\substack{n}}} \frac{n \langle {\bf u}_{n} \cdot {\bf u}_{n} \rangle}{\langle {\bf u} \cdot {\bf u} \rangle}.
\end{equation}
A notably smaller variation of $Ro_l \sim 0.5 \pm 0.1$ is found across the survey because the increase in flow speed ($Re$) with $Ro_\ell^M$ is largely offset by an increase in flow length scale ($\ell_U$). The velocity field is larger-scale in models with $Ro_\ell^M > 1$, where $\ell_U \sim 0.72$ on average, compared to those with $Ro_\ell^M \lesssim 1$ where $\ell_U \sim 0.26$.

\begin{figure*}
\center{
\noindent\includegraphics[width =36pc]{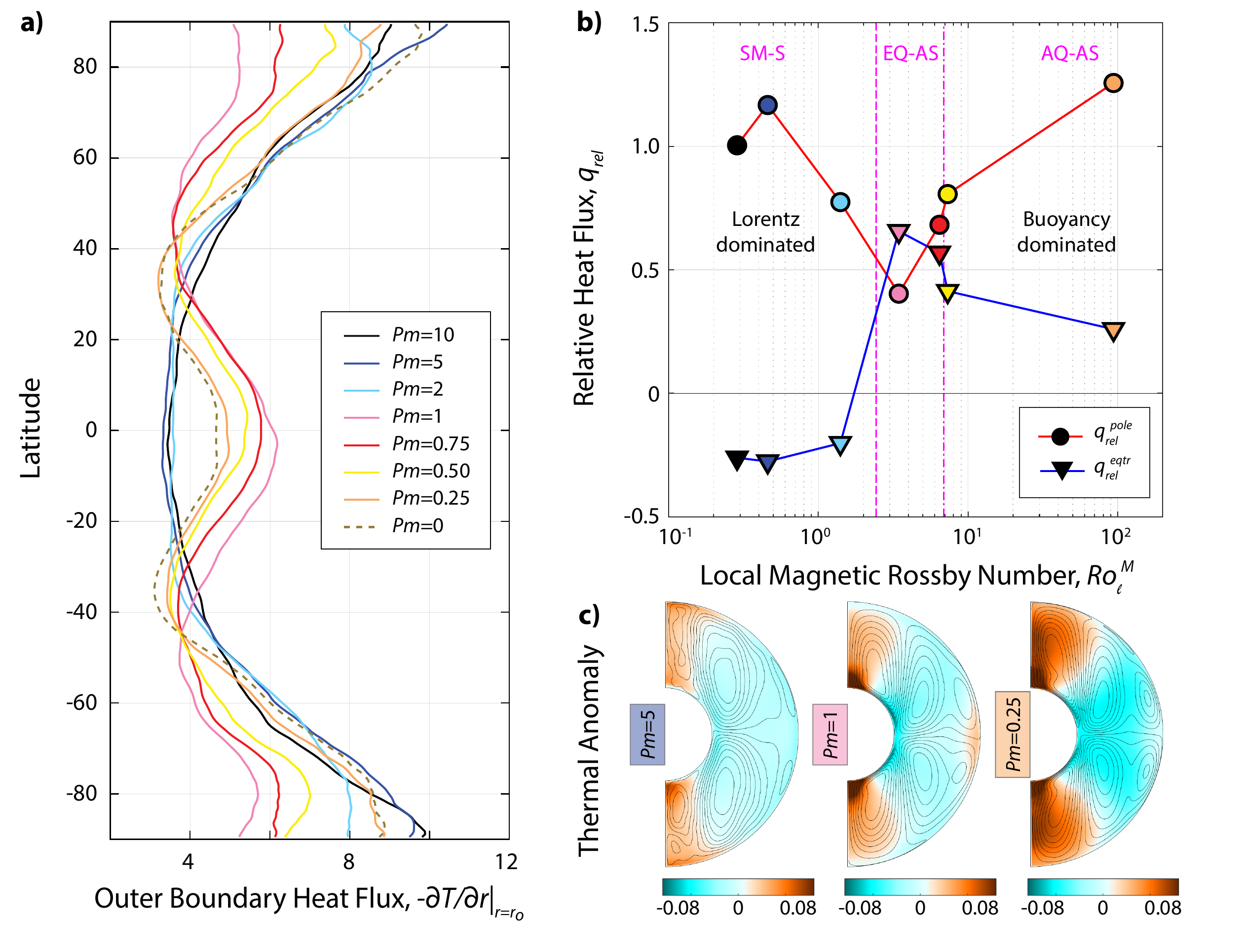}
\caption{Mean convective heat transfer characteristics, averaged over azimuth and time. {\bf (a)} Dimensionless heat flux on the outer shell boundary as a function of latitude. Color denotes $Pm$. {\bf (b)} Relative heat fluxes between the equator and mid-latitudes, $q^{eqtr}_{rel}$, and between the poles and mid-latitudes, $q^{pole}_{rel}$, plotted versus the local magnetic Rossby number. Vertical magenta lines demarcate approximate regime boundaries. {\bf (c)} Dimensionless thermal anomaly field (axisymmetric temperature field minus the spherically symmetric value) for the (left) $Pm=5$, (middle) $Pm=1$, and (right) $Pm=0.25$ cases. Red (blue) indicates the warmer (cooler) fluid, while solid (dashed) black contours denote clockwise (counterclockwise) circulations.}
 \label{fig:HF_Pm}}
 \end{figure*}

A fundamental change in zonal flow (i.e., differential rotation) also occurs in our fixed $Ro_C = 1.4$, varying $Ro_\ell^M$ survey. Figure~\ref{fig:ZF_Pm}a shows the time-averaged zonal flows on the outer shell boundary as a function of latitude, panel b shows the equatorial surface velocity as a function of $Ro_\ell^M$, and panel c compares zonal flows and meridional circulations for each regime. Two distinct styles of zonal flow are found. A strong retrograde equatorial jet and flanking prograde jets form when $Ro_\ell^M > 1$. Zonal flows in models with $Ro_\ell^M \lesssim 1$ are fundamentally different where, instead, a broad prograde equatorial jet dominates with retrograde flow near and interior to the tangent cylinder. In addition, the wind speeds in the prograde jet regime are only about half those in the retrograde jet regime. Since the dynamo fields and the associated Lorentz forces scale with electrical conductivity (i.e., $Pm$ and $(Ro_\ell^M)^{-1}$), the Figure~\ref{fig:ZF_Pm}a profiles demonstrate that magnetic field effects can flip the convection zone's differential rotation profile from solar-like to anti-solar states \citep[see also][]{Fan14}.

Figure~\ref{fig:ZF_Pm}b shows $\alpha_e$, the zonal velocity on the outer boundary at the equator as defined in equation~(\ref{eqn:alpha}), as a function of $Ro_\ell^M$. Here, we see that the flip in sign of $\alpha_e$ occurs for $Ro_\ell^M \simeq 1$, reminiscent of the change in zonal flow direction near $Ro_C \simeq 1$ shown in Figure~\ref{fig:MabGast}. As a result of these changes in differential rotation profile, we further define the dynamical regimes as SM-S, EQ-AS, AQ-AS to denote whether the equatorial zonal velocity fields are solar-like (S) or anti-solar (AS). 

Although the meridional circulations have similar trends in all of our models, important differences are evident in Figure~\ref{fig:ZF_Pm}c. Each hemisphere develops two large circulation cells with opposite polarities across the equator and within and outside of the tangent cylinder, which is the imaginary axial cylinder that circumscribes the inner shell boundary's equator \cite[e.g.,][]{AurnouEA03}. The circulation patterns differ, however, at large cylindrical radii. While smaller cells near the outer shell boundary occur in both regimes, their polarities are reversed. The cells reinforce the large-scale circulation pattern outside the tangent cylinder in the $Ro_\ell^M > 1$ models, but oppose it in the $Ro_\ell^M \lesssim 1$ models. This difference is likely important for the generation of zonal flows described above and for the pattern of convective heat transfer described below. 

\subsection{Convective Heat Transfer}
\label{sec:HF}

Table~\ref{tab:Pmruns} and Figure~\ref{fig:HF_Pm} show the characteristics of convective heat transfer in our models. Heat transfer efficiency is measured by the Nusselt number, which is the ratio of total to conductive heat flux across the shell:
\begin{equation}
    Nu = \frac{r_o}{r_i}\frac{qD}{\rho C_p \kappa \Delta T}
\end{equation}
where $q$ is heat flux per unit area on the outer shell boundary and $C_p$ is specific heat capacity. Since $\Delta T$ is fixed in our simulations, the conductive heat flux, which is proportional to $\Delta T / D$, is fixed. Changes in $Nu$ thus reflect variations in the amount of thermal energy transferred convectively across the shell. The Nusselt number ranges from $Nu = 12.6$ in the SM-S regime to $Nu = 13.5$ in the EQ-AS regime. Comparing against the non-magnetic value of $Nu = 13.0$, the dynamo can therefore act to either slightly diminish or slightly enhance the convective component of the heat transfer, depending on value of the local magnetic Rossby number.

Similar to the zonal flows, distinct styles of heat transfer profiles with latitude are found across our survey. Figure~\ref{fig:HF_Pm}a compares the outer boundary heat fluxes as a function of latitude for each model. For models in the SM-S regime with $Pm \geq 2$, heat transfer varies smoothly with latitude from a minimum at the equator to peaks at the poles. Conversely, for models in the EQ-AS and AQ-AS regimes, a secondary peak develops at low latitudes. This peak can be similar in amplitude to the poles in the EQ-AS regime and becomes less prominent as $Pm$ is decreased. 

Figure~\ref{fig:HF_Pm}b further quantifies this behavior as a function of local magnetic Rossby number through two relative heat flux parameters:
\begin{equation}
    q^{eqtr}_{rel} = \frac{( q_{eqtr} - \overline{q_{\pm 45^o}} )}{\overline{q_{\pm 45^o}}} \quad \text{and} \quad 
    q^{pole}_{rel} = \frac{( \overline{q_{poles}} - \overline{q_{\pm 45^o}} )}{\overline{q_{\pm 45^o}}},
    \label{eqn:qrel}
\end{equation}
where $q_{eqtr}$ is the heat flux at the equator, $\overline{q_{\pm 45^o}}$ is the average heat flux at $45^\circ$ latitude in the northern and southern hemispheres, and $\overline{q_{poles}}$ is the average heat flux at the northern and southern poles. Thus, $q^{pole}_{rel} > 0$ indicates that the polar heat flux exceeds the mid-latitude heat flux in all models \citep[cf.][]{Gastine23}. In contrast, $q^{eqtr}_{rel} < 0$ in the Lorentz dominated SM-S regime ($Ro_\ell^M \lesssim 1$) indicates that the equatorial region has lower flux than mid-latitudes, as expected for the low latitude minima shown in panel a. Interestingly, the highest $q^{eqtr}_{rel}$ and lowest $q^{pole}_{rel}$ values occur at $Ro_\ell^M \sim 3$. Thus, the commonly used $Pm=1$ value leads to an atypical behavior, where it is the only model to have higher equatorial than polar heat flux (i.e., $q^{pole}_{rel} - q^{eqtr}_{rel} < 0$).

Figure~\ref{fig:HF_Pm}c shows the thermal anomalies for models in each behavioral regime. All models have enhanced temperatures near the poles because strong upwellings develop here, regardless of the local magnetic Rossby number. The equatorial peaks in the models with anti-solar differential rotations ($Pm =$ 1, and 0.25 to a lesser degree) also correspond to regions of radially outward flow enhancement. In contrast, heat flux is minimal near the equator in the solar-like differential rotation model ($Pm=5$) because upwelling flows are locally weak at low latitudes. 

Significant differences in latitudinal heat flux are found in Figure~\ref{fig:HF_Pm}c between the equatorial, mid-latitude, and polar regions. Thus, latitudinal variations in observed heat flux may be diagnostic of the deep interior processes. Based on this, we argue that the collection of latitude-spanning heat flux data for stellar and solar objects can be of immense import towards understanding their internal MHD states \citep[e.g.,][]{hassler2023solaris}. Notably, the 2024–2033 Heliophysics Decadal Survey identifies the {\it Solar Polar Orbiter} as the highest-priority new mission within the Living With a Star program \citep{HelioDecadal}. 

\begin{figure*}
\centering{
\includegraphics[width =32pc]{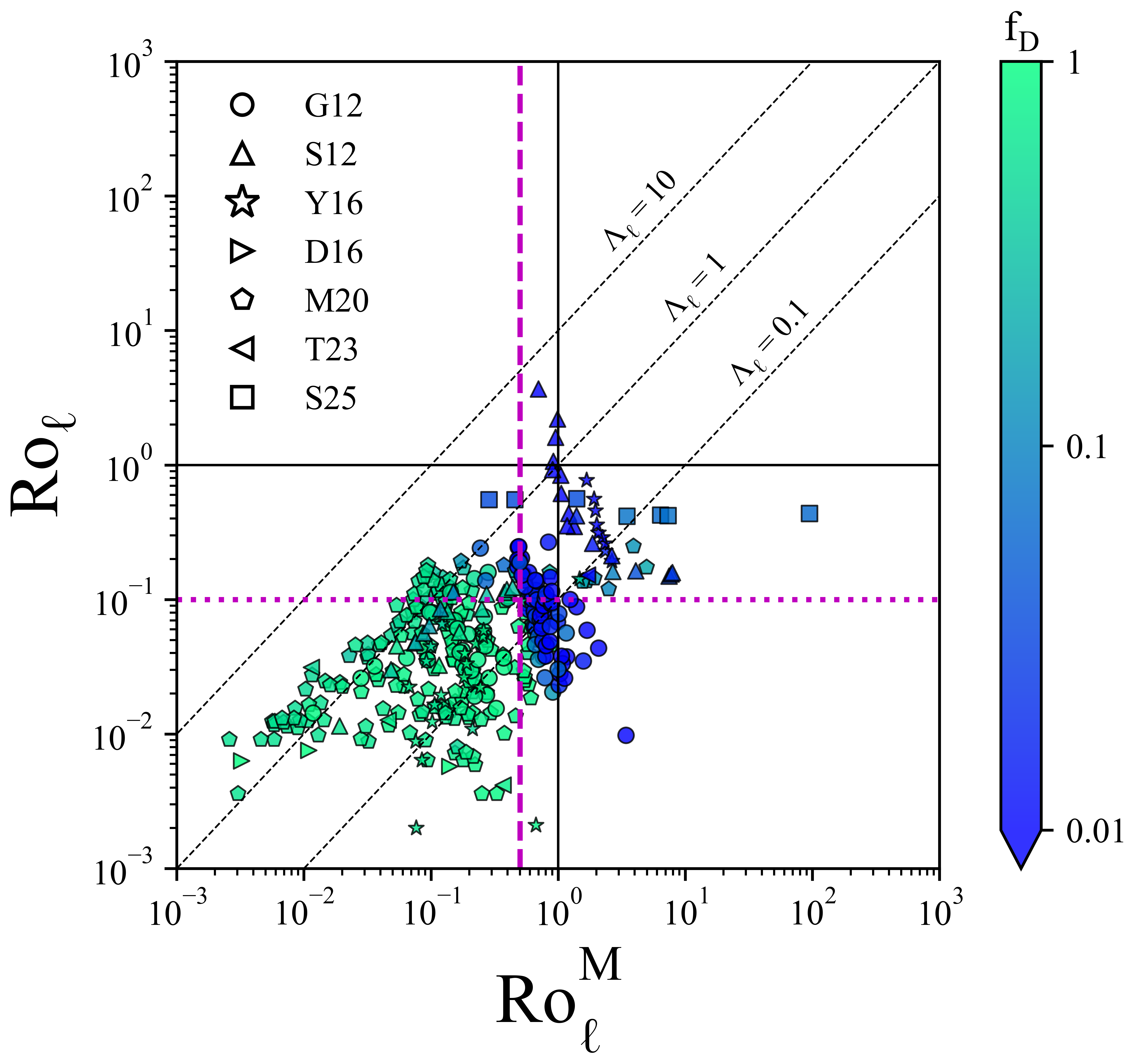}
\caption{Regime diagram showing dipolarity, $f_D$, in color as a function of the local Rossby number, $Ro_\ell$, and the local magnetic Rossby 
number, $Ro_\ell^M$. The purple lines indicate hypothesized transitions between dipole-dominated (green) and multipolar dynamos (blue): the vertical dashed line indicates $Ro_\ell^M = 0.5$ and the horizontal dotted line indicates $Ro_\ell = 0.1$. The thin dashed black lines denote constant $\Lambda_\ell$ values. Sources: \citet{GastineEA12} (G12), \citet{SoderlundEA12} (S12), \citet{dormy2016} (D16), \citet{Menu_20} (M20), \citet{Teed_Dormy_2023} (T23), and this study (S25).} 
\label{fig:Rol_RolM}}
\end{figure*}

\section{Discussion and Conclusions}

Here we have carried out a selected set of Boussinesq dynamo models in which we vary the value of the magnetic Prandtl number such that the intensity of the dynamo field varies between different cases, all made at fixed $Pr =1$ and $Ro_C = 1.4$. We purposely set $Ro_C$ near 1, just above the known hydrodynamical differential rotation transition point, in hopes of controlling the differential rotation transition by means of the Lorentz force. 

Relatively sharp transitions in differential rotation and dynamo morphologies are found in the vicinity of $Ro_\ell^M \sim 1$ in our models, accompanied by more modest changes in patterns of global heat transport. Thus, our simulations support the hypothesis that magnetic forces are a significant factor in the location of the behavioral transitions in turbulent MHD systems. 
Alternatively stated, it has been found that $Ro_\ell$ controls transitions in studies in which it is the broadly varied parameter \citep[e.g.,][]{Gilman78, AurnouEA07, GastineEA13, GuerreroEA13, KapylaEA14, MabuchiEA15, Camisassa22}, whereas $Ro_\ell^M$ is the transition parameter found here when it is the parameter that is broadly varied \cite[also see][]{Fan14,hotta2022generation,Menu_20, ZaireEA22,kapyla2023transition}. 

Our findings imply that the behavioral transitions in stellar convection zone dynamics are governed by a local-scale triple balance between inertia, Coriolis, and Lorentz forces. This idea is not new. For instance, \citet{CalkinsEA15} derived a set of quasi-geostrophic dynamo equations in which the local scale balance exists between the Coriolis, Lorentz, and buoyant inertial terms. This predicted local-scale triple balance has since been found to exist in a number of analyses of planetary dynamo simulations \cite[e.g.,][]{Yadav_16, SchwaigerEA19, aubert2020, schwaiger2021, nakagawa2022}. 

Figure~\ref{fig:Rol_RolM} further elucidates how convection zone regime transitions are controlled by a Coriolis, Lorentz, buoyant inertial triple balance. The dipolarity $f_D$ from an ensemble of stellar and planetary dynamo modeling studies is plotted as a function of $Ro_\ell^M$ on the abscissa and $Ro_\ell$ on the ordinate. Symbol shapes in Figure~\ref{fig:Rol_RolM} indicate the study from which the data were collected, while the symbol fill color represents the $f_D$ value, with a logarithmic scale transitioning from blue to green. The log color scaling pivots around $f_D \approx 0.1$. 
Dipolar $B$-field cases with $f_D \gtrsim 0.1$ have green fill colors; multipolar cases with $f_D \lesssim 0.1$ have blue fill. We believe this gives a sensible log-scale dipolarity cut-off given that $f_D$ varies by nearly six orders of magnitude in Figure \ref{fig:G12}. 

The two purple lines in Figure~\ref{fig:Rol_RolM} mark the estimated transition locations based on dipolarity data in Figure \ref{fig:G12}. The horizontal dotted purple line marks the postulated $Ro_\ell \simeq 0.1$ transition \cite[e.g.,][cf.~Figure \ref{fig:G12}c]{Christensen06}, while the vertical dashed purple line marks the estimated $Ro_\ell^M \simeq 0.5$ dipolarity transition in Figure \ref{fig:G12}d. In addition, the thin black dashed diagonals in Figure~\ref{fig:Rol_RolM} correspond to constant values of $\Lambda_\ell$ since, following (\ref{eq:RoMell}), the local Elsasser number is defined as $\Lambda_\ell = Ro_\ell/Ro_\ell^M$. 

The data compilation in Figure~\ref{fig:Rol_RolM} comes from a broad array of spherical shell dynamo modeling studies, both stellar and planetary, anelastic and Boussinesq, and over a range of $Pr$, $Pm$ and thermo-mechanical boundary conditions. Despite the broad range of sources, robust trends exist in the $f_D$ data, defining different behavioral regimes of dynamo generation in this $(Ro_{\ell}^M, Ro_\ell)$ parameter space. First, dipolar cases ($f_D \gtrsim 0.1$) exist primarily below $Ro_\ell = 0.2$ and $Ro_\ell^M = 0.5$. 

The dipolar cases are not evenly spread across the $(Ro_\ell^M < 0.5, \ Ro_\ell < 0.2)$ quadrant of Figure~\ref{fig:Rol_RolM}. Instead, they are bounded by the $\Lambda_\ell \approx 3$ diagonal line from above. Thus, the dipolar cases occupy a wedge-like region bounded by $(Ro_\ell^M < 0.5, \Lambda_\ell < 3)$. 
An implication of this wedge structure is that the $Ro_\ell \sim 0.1$ transition in dynamo morphology proposed in \citet{ChristensenAubert06} may hold only locally in the vicinity of $Ro_\ell^M \approx 0.5$ in Figure \ref{fig:Rol_RolM}, where the $\Lambda_\ell \sim 1$ and $Ro_\ell^M \sim 0.5$ lines roughly intersect. If only multipolar dynamo cases are found to exist in the $Ro_\ell \lesssim 0.1$ and $\Lambda_\ell \gtrsim 3$ region of parameter space, then the hydrodynamic $Ro_\ell \approx 0.1$ dynamo transition will need to be replaced with an MHD argument that better fits the data.

The fixed $Ro_C = 1.4$ data from this study are demarcated by square symbols in Figure \ref{fig:Rol_RolM}. With $Ro_\ell$ values ranging from 0.4 to 0.6, our cases all lie above the dipolar wedge and, indeed, all have nondipolar magnetic fields. Yet changes in magnetic field morphology (Table \ref{tab:regimes}) and differential rotation pattern (Figure~\ref{fig:ZF_Pm}) still occur in our cases, occurring over $Ro_\ell^M \approx 2$ to 7. This suggests that $Ro_\ell^M = \mathcal{O}(1)$ may control the behavior of convection zone transitions over a wide range of $ Ro_\ell$ values, possibly including $Ro_\ell \gg 1$.

The dipolar wedge's $\Lambda_\ell \approx 3$ bounding line in Figure \ref{fig:Rol_RolM} suggests that dipolar dynamo simulations are attracted to convection-scale dynamics that are in local magnetostrophic balance with $\Lambda_\ell \sim 1$ \citep[e.g.,][]{CalkinsEA15, KingAurnou15, Yadav_16, SoderlundEA15, Aurnou17, schwaiger2021, hotta2022generation}. The $Ro_\ell^M \approx 0.5$ upper bound on the wedge implies that dipolar solutions are stable until inertial accelerations become comparable to the Lorentz terms, in good agreement with the force balance calculations of \citet{Yadav_16} and following studies. Thus, the wedge suggests that dipolar dynamo action in current-day simulations exists in a triple balance where local-scale Coriolis forces may be approached by Lorentz forces from below, which may be approached by inertial forces also from below. 

Careful inspection of Figure \ref{fig:Rol_RolM} suggests that the separation between the dipolar and multipolar dynamo morphology cases occurs along a line that is slightly off-vertical. Determining the robustness of this tilted transition line and elucidating the physics that sets its slope are open topics. Further, the behavior of dynamos in the high $Ro_\ell$, high $Ro_\ell^M$ region of parameter space remains open as well. It also remains to be shown how best to replace the \textit{a posteriori} local Rossby numbers with \textit{a priori} control parameters. A version of Figure~\ref{fig:Rol_RolM} made with control parameters on both axes will lead to far more predictive power and more meaningful testing of regime transition hypotheses. 

A comparison of Figures~\ref{fig:Bmorph} and \ref{fig:ZF_Pm} suggests that behavioral transitions in differential rotation and dynamo morphology may be co-located. 
To better address this, Figure~\ref{fig:Rol_RolM} should ideally include a second panel highlighting transitions in differential rotation, based on an ensemble of $\alpha_e$($Ro_\ell^M, \ Ro_\ell$) values from spherical shell dynamo simulations. 
To our consternation, a comparable $\alpha_e (Ro_\ell^M, \ Ro_\ell)$ plot remains to be made, since we were unable to compile an ensemble $\alpha_e$ dataset from the existing literature. 
Thus, we must leave unanswered whether or not broadly similar regime boundaries exist for transitions between solar and anti-solar differential rotation as we have found for dynamo morphology in Figure~\ref{fig:Rol_RolM}. 

The goal of this study has been to show that the regimes of spherical shell dynamo physics are controlled by a local-scale, multi-term MHD balance dominated by the Coriolis, Lorentz, and inertial terms. This has been done here via a relatively simplistic set of Boussinesq simulations. In contrast, stellar convection zones are strongly density-stratified. Although the transition between anti-solar and solar-like differential rotation states seems to be insensitive to density stratification \citep[e.g.,][]{GastineEA14_MNRAS}, the details of the differential rotation profile are affected \citep[e.g.,][]{KMB11}. This is of importance in order to understand how, for example, the solar differential rotation profile arises. Similarly, the large-scale dynamo solutions are also affected by density stratification, such that multipolar solutions tend to be favored over dipolar ones at sufficiently strong stratification \citep[e.g.,][]{GastineEA12}. The propagation direction of dynamo waves, again relevant for the solar case, is another example where stratification can play a role \citep[e.g.,][]{KMCWB13}. 

Furthermore, a small-scale dynamo that produces fields around the scale of the turbulence and which also appears in the absence of flow helicity and differential rotation is excited at sufficiently high $Rm$ \citep[e.g.,][]{hotta2022generation,kapyla2023transition,2024arXiv240608967W}. Another aspect that can be improved upon in comparison to the current simulations is that in stellar convection zones, the thermal Prandtl number is $Pr\ll 1$. Dynamics of convection at realistic $Pr$ are likely to be quite different from those near $Pr=1$ where most numerical dynamo simulations are made \citep[e.g.,][]{CBTMH91,2021A&A...655A..78K, KingAurnou13, SchaefferEA17, GuervillyEA19, Grannan2022, xu2022, horn2024}. Isolating the effects of density stratification, small-scale dynamo action, and more realistic Prandtl numbers on the large-scale magnetic, velocity, and thermal
fields all warrant future systematic studies.

\section*{Acknowledgements}

The authors thank E.~Dormy, T.~Gastine, R.~Teed, and B.~Zaire for providing datasets, along with T.~Schwaiger for helpful discussions. The authors would like to thank the Isaac Newton Institute for Mathematical Sciences, Cambridge, UK, for support and hospitality during the programme ``Dynamos in planets and stars - similarities and differences", supported by EPSRC grant EP/R014604/1, where work on this paper was undertaken. KMS additionally acknowledges support from the National Science Foundation (NSF) Astronomy and Astrophysics program via award AST \#2308185. PNW acknowledges support from the NASA Juno project and NASA CDAP grant No. 80NSSC23K0511. PJK acknowledges support by the Deutsche Forschungsgemeinschaft (DFG, German Research Foundation) Heisenberg programme (grant No. KA 4825/4-1), and by the Munich Institute for Astro-, Particle and BioPhysics (MIAPbP) which is funded by the DFG under Germany’s Excellence Strategy – EXC-2094 – 390783311. JMA thanks the NSF Geophysics Program for support via award EAR \#2143939. Computing resources supporting this work were provided by the NASA High-End Computing (HEC) Program through the NASA Advanced Supercomputing (NAS) Division at Ames Research Center. We have no conflicts of interest to disclose. 

\section*{Data Availability}

We use the dynamo code MagIC \citep{Wicht02,GastineWicht12} with the SHTns library to efficiently calculate the spherical harmonic transforms \citep{Schaeffer13}; the open source code is available at \href{https://github.com/magic-sph/magic}{https://github.com/magic-sph/magic}. Model results are tabulated in the manuscript. 


\bibliographystyle{mnras}
\bibliography{biblio_combined_Pm} 




\appendix
\section{Definitions in previous Studies}

\begin{table*}
\setlength{\tabcolsep}{3pt}
\renewcommand{\arraystretch}{1.2}
\centering 
\begin{tabular}{l|l|l|ll}
\hline 
Publication & Velocity length scale in $Ro_\ell$ & Magnetic length scale in $\Lambda_\ell$ & \multicolumn{2}{c}{$f_D$} \\ 
 &  &  & num. & denom. \\ 
\hline  
G12 & $\pi/\overline{n}_U$ & $\pi/\left(2 \sqrt{\overline{n}_B^2 + \overline{m}_B^2}\right)$ & $\mathcal{P}(n=1,m=0)$ & $\Sigma_{n=1}^{n_{max}} \mathcal{P(\mbox{\it n})}$ \\
S12 & $\pi/\left(\sqrt{\overline{n}_U^2 + \overline{m}_U^2}\right)$ & $\pi/\left(2 \sqrt{\overline{n}_B^2 + \overline{m}_B^2}\right)$ & $\mathcal{P}(n=1)$ & $\Sigma_{n=1}^{n_{max}} \mathcal{P(\mbox{\it n})}$ \\
Y16 & $\pi/\overline{n}_U$ & $\pi/(2 \overline{n}_B)$ & $\mathcal{P}(n=1,m=0)$ & $\Sigma_{n=1}^{n_{max}} \mathcal{P(\mbox{\it n})}$ \\
D16 & $\pi/\overline{n}_U$ & $\sqrt{\int_V\boldsymbol{B}^2dV \,/\int_V(\nabla\times\boldsymbol{B})^2dV}$ & $\mathcal{P}(n=1,m=0)$ & $\Sigma_{n=1}^{n=12} \mathcal{P(\mbox{\it n})}$ \\
M20 & $\pi/\overline{n}_U$ & $\sqrt{\int_V\boldsymbol{B}^2dV \,/\int_V(\nabla\times\boldsymbol{B})^2dV}$ & $\mathcal{P}(n=1)$ & $\Sigma_{n=1}^{n=12} \mathcal{P(\mbox{\it n})}$ \\
T23 & $\pi/\overline{n}_U$ & $\sqrt{\int_V\boldsymbol{B}^2dV \,/\int_V(\nabla\times\boldsymbol{B})^2dV}$ & $\mathcal{P}(n=1,m=0)$ & $\Sigma_{n=1}^{n=12} \mathcal{P(\mbox{\it n})}$ \\
S25 (this study) & $\pi / \overline{n}_{U}$ & $\pi/\left(2 \sqrt{\overline{n}_B^2 + \overline{m}_B^2}\right)$ & $\mathcal{P}(n=1)$ & $\Sigma_{n=1}^{n=10} \mathcal{P(\mbox{\it n})}$ \\  
\hline
\end{tabular}
\caption{Length scales used to calculate $Ro_\ell$ and $\Lambda_\ell$, and definition of dipolarity $f_D$, split into numerator (indicating total dipole or axial dipole power) and denominator (relative to the power in the spectrum up to degree $n=12$ or model resolution). See equations~(\ref{eq:magl}) and (\ref{eq:vell}) for definitions of $\overline{n}_B$, $\overline{m}_B$ and $\overline{n}_U$, with $\overline{m}_U$ being the azimuthal equivalent of $\overline{n}_U$.}
\label{tab:definitions}
\end{table*}



\bsp	
\label{lastpage}
\end{document}